\documentstyle[preprint,aps]{revtex}
\tighten
\draft
\begin{document}
\title{ Orthogonality catastrophe in a mesoscopic conductor
due to a time-dependent flux.}
\author{Hyunwoo Lee, L.S.Levitov\cite{*}}
\address{Massachusetts Institute of Technology,
12-112, Department of Physics,\\
77 Massachusetts Ave., Cambridge, MA 02139}
\maketitle

\begin{abstract}
We consider quantum fluctuations of current in a metallic loop
induced by varying magnetic flux. The dependence of the
fluctuations on the flux change $\Phi$ contains a
logarithmically divergent term periodic in $\Phi$ with the
period $\Phi_0=hc/e$. The fluctuation is smallest near
$\Phi=n\Phi_0$. The divergence is explained by a comparison with
the orthogonality catastrophe problem. The $\Phi_0-$periodicity
is related with the discreteness of "attempts" in the binomial
statistics picture of charge fluctuations.

\end{abstract}
\pacs{PACS numbers: 72.10.Bg, 73.50.Fq, 73.50.Td}
\narrowtext

The orthogonality catastrophe problem emerged from the
observation that the ground state of a Fermi system with a
localized perturbation is orthogonal to the non-perturbed ground
state, no matter how weak the
perturbation\cite{Anderson}. Originally, the discussion was
focused on the purely static effect of Fermi correlations on the
ground state that leads to the orthogonality, but then it
shifted to dynamical effects. When a sudden localized
perturbation is switched on in a Fermi gas, the number of excited
particle-hole pairs detected over a large time interval $t$
diverges as $\ln t/\tau$, where $\tau$ is the time of switching
of the perturbation. This effect leads to power law
singularities in transition rates of processes that involve
collective response of fermions, e.g., in X-ray absorption in
metals\cite{Nozieres,Mahan}. In this paper we present an
application of the orthogonality catastrophe picture to the
theory of quantum noise in electric circuits.

Due to Callen-Welton's theorem, the fluctuations in equilibrium
are related in a universal way with kinetic response \cite{1}.
However, due to the lack of such relation in the non-equilibrium
situation, physics of the non-equilibrium noise may be quite
different from that of corresponding kinetic response. In this
paper, we study noise in a conductor driven from one equilibrium state
to another due to changing flux by a constant amount over fixed time,
and find that charge fluctuations distinguish between integer
and non-integer flux change in a way resembling Aharonov-Bohm
effect. Let us emphasize that the geometry of our conducting
loop excludes the usual dc Aharonov-Bohm effect.

Noise in quantum conductors received an increased attention
recently due to its relevance for applications and also because
it leads to understanding interesting effects in quantum
statistics. Initially, only the spectral density of the noise was
considered\cite{2,3,4,5,6}; however, it was realized shortly that
it is also quite useful to have the picture of charge
fluctuations in the time domain, since in the spectral density
many important features remain implicit\cite{7}. In particular,
the description of the fluctuations in the time domain allows us to
approach microscopically the problem of the distribution of
charge transmitted through the system over fixed time. This
distribution is binomial, i.e., it results from Bernoulli
statistics\cite{8}. We discuss below a relation of the binomial
distribution with the contribution to the noise periodic in
flux.

Let us consider a metallic conductor bent in a loop of length $L$
threaded by a time-dependent magnetic flux (Fig.~\ref{figure1}).
Ends of the conductor are treated as ideal leads that serve as
infinite reservoirs with equilibrium distribution. The flux
increases over a time $2\tau$ from zero to a constant value:
$\Phi(t\ll-\tau)=0$, $\Phi(t\gg\tau)=\Phi$. It is assumed that
the diffusion picture is valid, i.e., $L\gg l$, the mean free
path, and that the time $\tau\gg t_f= 3L^2/v_Fl$, the duration of
diffusion across the loop. We
are interested in fluctuations of the charge $Q$ transmitted
through the conductor over a much longer time interval
$-t_0<t<t_0$, $t_0\gg\tau$. For the transmitted charge
$Q=\int_{-t_0}^{t_0}j(t)dt$, we get the expected Ohm's law for
the average $\langle Q\rangle= {\rm g} eG_0 \Phi/\Phi_0$, where
$g$ is spin degeneracy and $G_0$ is dimensionless conductance
expressed through transmission coefficients:
$G_0=\sum_nT_n$, in accordance with Landauer's formula\cite{9}.
For the mean square of the charge fluctuation in the limit of
low temperature $T\ll\hbar/t_0$ and $\Phi\gg\Phi_0$, we obtain
   \begin{equation}\label{result}
\langle\!\langle Q^2\rangle\!\rangle=\ {\rm g} e^2
G_1\ [{2\over\pi^2}\sin^2{\pi\Phi/ \Phi_0} \ln {t_0/\tau}\ + \
{\Phi/\Phi_0}]\ +\ \dots\ , \end{equation}
   where $G_1=\sum_n T_n(1-T_n)$. The dots in (\ref{result})
represent correction higher order in $\Phi_0/\Phi$ and the
equilibrium noise
   \begin{equation}\label{equil}\langle\!\langle
Q^2\rangle\!\rangle_{eq}=\ {{\rm g} e^2 G_0\over\pi^2}\ \ln
t_0/t_f \end{equation}
   that follows from the Nyquist formula $\langle\!\langle
j_\omega j_{-\omega}\rangle\!\rangle={\rm g} e^2 G_0
\omega{\rm coth}\omega/2T$ taken at $T=0$, Fourier transformed and
combined with the relation $Q=\int_{-t_0}^{t_0}j(t)dt$.

In order to make obvious the mapping on the orthogonality
catastrophe problem, let us consider an ideal single channel
conductor, i.e., the Schr\"odinger equation
  \[i{\partial\over\partial t}\psi (x,t)= [{1\over
2}(-i{\partial\over\partial x }-{e\over c}A(x,t))^2+U(x)]\psi
(x,t) \]
   in one dimension, where the potential $U(x)$ represents the
scattering region and $A(x,t)$ is the vector potential
corresponding to the flux $\Phi(t)$. Since the switching time
$\tau$ is assumed to be much longer than the transport time
$t_f$, one can treat the vector potential as static and apply a
gauge transformation in order to accumulate the flux $\Phi(t)$
in the phases of the transmission amplitudes, thus making them
time dependent:
   \begin{equation}\label{ampl}
A_{L(R)}(t)=A_{L(R)}\ e^{\pm i\phi(t)}\ ,\end{equation}
   where $\phi(t)= 2\pi \Phi(t)/\Phi_0$. Then one can write the
scattering states with energy close to $E_F$ as
    \[{\psi}_{L,k}(x,t)=e^{-iE_kt}\left\{\begin{array}{ll}
e^{ikx} +B_L\ e^{-ikx}, & x<-L/2 \\
A_L(t_r)e^{ikx}, & x>L/2 \end{array}
\right. ,\]
    \begin{equation}\label{scatt}
{\psi}_{R,k}(x,t)=e^{-iE_kt}\left\{\begin{array}{ll}
A_R(t_r)e^{-ikx}, & x<-L/2 \\
e^{-ikx}+B_R\ e^{ikx}, & x>L/2 \end{array}
\right. ,
 \end{equation}
where the retarded $t_r=t-|x|/v_F$ takes care of the finite
speed of motion after scattering. The effect of the flux on the
scattering phases $\delta_1$, $\delta_2$ can be found by
diagonalizing the scattering matrix
  \begin{equation}\label{matr} \hat{\cal S}(t)=\left[\matrix{
  A_Le^{i\phi(t)} & B_R \cr
  B_L & A_Re^{-i\phi(t)} \cr }\right]
  \end{equation}
that has eigenvalues $e^{i\delta_1},e^{i\delta_2}$. The relation
between the phases $\delta_{1,2}$ at $t\ll -\tau$ and their
values $\delta_{1,2}'$ at the moment $t$ is written conveniently
through $\delta_\pm =(\delta_1\pm \delta_2)/2$:
   \begin{equation}\label{delta} \cos^2\delta_-'+\cos^2\delta_{-}
-2\cos\delta_-'\cos\delta_-\cos \phi(t) =|A_L|^2\sin^2\phi(t)\
.\end{equation}
    Now the situation can be readily understood in terms of the
orthogonality catastrophe in the Fermi system with the time
dependent perturbation (\ref{matr}). Change of the flux induces
the shift of the phases $\delta_\pm\rightarrow\delta_\pm'$ and
makes the new ground state orthogonal to the old one:
    \begin{equation}\label{orth} \langle 0' | 0\rangle =
\exp[-2{\delta^2\over\pi^2}\ln E_F/\Delta ]\ , \end{equation}
   where and $\Delta$ is level spacing near $E_F$and
$e^{i\delta}$ is an eigenvalue of the matrix $\hat{\cal
S}^{-1}(t)\hat{\cal S}(t\ll -\tau)$:
$\sin\delta/2=|A_L|\sin\phi(t)/2$. In terms of dynamics this
implies that the old ground state evolves to a state with
infinitely many excited particle-hole pairs, in agreement with
the standard orthogonality catastrophe calculation\cite{Mahan}.
We shall see below that this leads to a logarithmically
diverging contribution to noise, since for each of the
particle-hole pairs there is a finite probability (equal to
$|A_LB_R|^2$) that the particle and the hole will go to
different ends of the loop, thus resulting in a current
fluctuation. The periodicity in $\Phi$ follows from the gauge
invariance and is explicit in Expr.(\ref{delta}) for
$\delta_\pm'$. The logarithmic divergence vanishes at
$\Phi=n\Phi_0$, as should be expected, since at integer $\Phi$
there is no long term change of the scattering.

Second term in (\ref{result}) is interesting in connection with
the picture of binomial statistics for the noise at low
temperature \cite{8}. In the case of dc bias, the distribution of
charge for a single channel situation was found to be binomial
with frequency of attempts equal to $eV/h$ and the probabilities
of outcomes $p=D$, $q=1-D$, $D=|A_L|^2$. If understood
literally, this means that the attempts to transfer charge are
repeated regularly in time, almost periodic with the period
$h/eV$, with each attempt having two outcomes -- transmission or
reflection -- occurring with the probabilities $p$ and $q$.
However, the regularity of the attempts does not lead to an ac
component of current, rather it appears just as a part of statistical
description of charge fluctuations. Still, the presence of a
non-zero frequency in a non-interacting system requires
interpretation.

Let us suppose that the flux varies linearly with time,
$\Phi(t)=-cVt$. Because of the expression
$-\partial\Phi/c\partial t$ for the e.m.f., the linear dependence
of $\Phi(t)$ in its effect on the noise is equivalent to
constant voltage $V$. In accordance with one's expectation,
second term of Expr.(\ref{result}) for a single channel is
$\langle\!\langle Q^2\rangle\!\rangle=\ {\rm g} e^2D(1-D)
\Phi/\Phi_0$, i.e., it is precisely of the form arising from the
binomial distribution with probabilities of outcomes $p=D$ and
$q=1-D$ and the number of attempts $N=\Phi/\Phi_0$. (Let us
recall that the second moment of the binomial distribution equals
$pqN$.) Taking into account that the time during which the flux
changes by $\Phi_0$ is $h/eV$, we can interpret the number of
attempts in the statistical picture as the number of flux quanta
by which the flux is changed. Such a conclusion suggests an
interesting generalization of the picture of binomial statistics
by attributing the meaning of the number of attempts to the flux
change measured in units of $\Phi_0$, regardless of the linear or
non-linear character of the flux dependence on time.

Although this picture is yet to be confirmed by analytic
treatment, let us remark that it receives some support from the
property of the $\Phi_0-$periodic term in (\ref{result}) to
vanish at integer $\Phi$. One may conjecture that the statistics
are purely binomial only when the flux change is an integer and has
diverging logarithmic corrections otherwise. The distinction
that Expr.(\ref{result}) makes between integer and non-integer
values of the flux and the relation of integer flux change to
the number of attempts in the binomial distribution, gives
another perspective to the statistical picture of charge
fluctuations.

Now let us turn to the calculation. Since the characteristic
times we have are longer than $t_f=3L^2/v_Fl$, we can find the
fluctuations in one channel and then sum over all conducting
channels. Let us start with the time-dependent scattering states
(\ref{scatt}) and write operators of second quantized electrons
as $\hat\psi(x,t)= \hat\psi_L(x,t)+ \hat\psi_R(x,t)$,
$\hat\psi_{L(R)}(x,t)= \sum_k \psi_{L(R),k}(x,t) \hat
a_{L(R),k}$. Then we write current operator $\hat
j=-i{e\hbar\over m}\hat\psi^+\nabla\hat\psi$ and by averaging
$\hat Q=\int_{-t_0}^{t_0}\hat j(t)dt$ obtain
$\langle Q\rangle={\rm g}eD\Phi/\Phi_0$, consistent with the Ohm's law.
Then we compute current-current correlation function
$\langle\!\langle\hat j(t_1)\hat j(t_2)\rangle\!\rangle=$
    \begin{equation}\label{j-j}
{e^2\over h^2}\sum_{E,E'}e^{-i(E-E')(t_1-t_2)} {\big [}
|A_1A_2|^2f^1_{E,E'} +\bar B_1A_1 \bar A_2B_2 f_{E,E'}+
\bar A_1B_1 \bar B_2A_2
f_{E',E}{\big ]} ,\end{equation} where $A_{1,2}=A_L(t_{1,2})$,
$B_{1,2}=B_R(t_{1,2})$, $f_{E,E'}=n(E')(1-n(E))$, $f^1_{E,E'}=
f_{E,E'}+f_{E',E}$.
 To obtain the second moment of charge, we integrate over $-t_0 <
t_{1,2} < t_0$, then substitute time dependent scattering
amplitudes (\ref{ampl}) and $n(E)= (e^{E/T}+1)^{-1}$. The
result reads
    \begin{equation}\label{1}\langle\!\langle
Q^2\rangle\!\rangle= {{\rm g}e^2\over2\pi}\int {\Big [} D^2
{\Big |}\int_{-t_0}^{t_0}e^{i\omega t}dt{\Big |}^2 +\ D(1-D)\
{\Big |} \int_{-t_0}^{t_0}e^{i\phi(t)+i\omega t}dt{\Big |}^2
{\Big ]}\ \omega{\rm coth}{\hbar\omega\over 2T}\ {d\omega\over
2\pi}\ ,\end{equation} where $\phi(t)=2\pi\Phi(t)/\Phi_0$.

The first term in (\ref{1}) is a part of equilibrium noise since it
does not depend on $\Phi$. To analyze the second term, we first look
at the step-like time dependence, $\phi(t<0)=0,\
\phi(t>0)=2\pi\Phi/\Phi_0$, in order to get terms diverging as
$t_0\rightarrow\infty$ and $T=0$. Thus, we have
   \begin{equation}\label{2} {{\rm g}e^2\over2\pi} \int {\Big
|}{1-e^{-i\omega t_0}\over i\omega} +e^{2\pi
i\Phi/\Phi_0}{e^{i\omega t_0}-1\over i\omega}{\Big |}^2
|\omega|{d\omega\over2\pi} = {{\rm g}e^2\over\pi^2}
(1+2\sin^2\pi\Phi/\Phi_0) \ln {t_0\over t^*}\ .\end{equation}
  Logarithmic contribution to the non-equilibrium noise
(\ref{result}) is obtained by subtracting the result for
$\Phi=0$ as corresponding to equilibrium, and then substituting
cutoff time $t^*=\tau$.

The term of Expr.(\ref{result}) proportional to $\Phi/\Phi_0$ is
obtained by rewriting the second term of (\ref{1}) as
    \begin{equation}\label{3} {{\rm g}e^2\over2\pi}D(1-D)\int
{d\omega\over 2\pi} |\omega|
\int dt_1\int dt_2 e^{i\phi(t_1)-i\phi(t_2)+i\omega(t_1-t_2)}\
\end{equation}
  and extracting the contribution of almost coinciding times
$t_1$ and $t_2$ by going to new variables $t=(t_1+t_2)/2$,
$t'=t_1-t_2$ and changing order of integrations:
   \begin{equation}\label{4}\int dt\int {d\omega\over 2\pi}
|\omega|\int dt' e^{i\phi(t+t'/2)-i\phi(t-t'/2)+i\omega t'}=\
\int|{\dot\phi}|dt \ ,\end{equation}
  where we replaced $\phi(t+t'/2)-\phi(t-t'/2)$ by ${\dot\phi}\
t'$. The result (\ref{4}) is approximate: it does not give the
$log$ term because the transformation (\ref{4}) properly takes
care of the integral (\ref{3}) only in the domain $t_1\simeq
t_2$, under the restriction that $\Phi(t)$ is varying
sufficiently smoothly. When $\Phi(t)$ is a monotonous function,
${\dot\phi}>0$, the integral in the r.h.s. of (\ref{4}) equals
$2\pi\Phi/\Phi_0$ and thus produces the second term of
Expr.(\ref{result}).

It is clear from the derivation that the two terms of
Expr.(\ref{result}) arise from different integration domains in
the $t_1-t_2$ space: the first term corresponds to
$|t_{1,2}|\ge\tau, \ t_1t_2<0$, while the second one is due to
almost coinciding moments, $|t_1-t_2|\ll\tau$. Since the domains
are almost non-overlapping, the two contributions to the noise
(\ref{result}) do not interfere (cross terms are small).

In order to have a feeling of how big the correction to
Expr.(\ref{result}) can be, we shall study two examples. Firstly,
let us consider functions
$\Phi_\alpha(t)={\alpha\over\pi}\Phi_0\tan^{-1}t/\tau$
corresponding to
$e^{i\phi_\alpha(t)}=((1+it/\tau)/(1-it/\tau))^\alpha$,
$\Phi=\alpha\Phi_0$. For integer $\alpha$ Expr.(\ref{3}) can be
evaluated exactly using technique of generating functions. The
one useful here is
   \begin{equation}\label{gener}{\rm
F}_\omega(z)=\sum_{n>0}z^n\int e^{i\omega t+i\phi_n(t)}dt=
\left\{\begin{array} {ll} 0, & \omega>0 \\ {4\pi
z\over(1+z)^2}e^{\omega(1-z)/(1+z)}, & \omega<0 \end{array}
\right.\ .\end{equation}
     The last equality is a result of changing
order of summation and integration. Then it is straightforward
to compute
    \begin{equation}\label{5}\int|\omega|{\rm
F}_\omega(z_1)\overline{{\rm F}_\omega(z_2)}{d\omega\over2\pi}=
2\pi{z_1\bar z_2\over (1-z_1\bar z_2)^2}\ .\end{equation}
  L.h.s. of (\ref{5}) expanded in series generates integrals
(\ref{3}) for $\phi(t)=\phi_n(t)$ as coefficients of the terms
$z_1^n\bar z_2^n$. On the other hand, r.h.s. of (\ref{5}) equals
$\sum_{n>0}2\pi nz_1^n\bar z_2^n$. This means that for all
$\Phi_n(t)$ Expr.(\ref{result}) is exact. With this result, we
can state that for arbitrary $\alpha$ inaccuracy of
Expr.(\ref{result}) is at most a bounded function of $\Phi$
taking zero value at every integer $\Phi/\Phi_0$.

The next function we consider rises linearly in the interval
$-\tau<t<\tau$ and is constant for $|t|>\tau$. In this case
  \begin{equation}\label{6}\int_{-t_0}^{t_0}e^{i\phi(t)+i\omega
t}dt={2\over\omega} e^{\pi i\Phi/\Phi_0}[\sin\omega
t_0-{\pi\Phi\over\omega\tau\Phi_0+\pi\Phi}
\sin(\omega\tau+\pi\Phi/\Phi_0)]\ .\end{equation}
  After substituting this in (\ref{1}) and subtracting the
equilibrium part not vanishing at $\Phi=0$ one gets
  \begin{equation}\label{7} {\rm g}e^2D(1-D)\int {\Phi^2
\sin^2(\omega\tau+\pi\Phi/\Phi_0) \over
(\omega\tau\Phi_0+\pi\Phi)^2 |\omega|} d\omega \ .\end{equation}
  The integral is converging for $\Phi=n\Phi_0$ and
logarithmically diverging otherwise, as it should be. For large
integer $n$ one easily evaluates it using the  expansion
  \begin{equation}\label{10}\int_{-\infty}^\infty
{\sin^2x\over(x+\pi n)^2}{dx\over |x|} ={1\over n}+{\ln
n\over\pi^2n^2}+{\rm O}(n^{-2})\ ,\end{equation} and obtains
correction to Expr.(\ref{result}):
  \begin{equation}\label{correction}{{\rm
g}e^2\over\pi^2}D(1-D)\ln\Phi/\Phi_0 \end{equation} plus higher
order terms. As one might have expected, the noise comes to be
stronger for our second function since it is less smooth.
However, the correction (\ref{9}) is relatively
small, even for it.

At this point let us recall that in order to generalize our
results for a single channel
to the case of a mesoscopic conductor containing
many conducting channels, one
just needs to replace $D(1-D)$ by $\sum_nT_n(1-T_n)$. The
condition of validity of our treatment then is $\tau\gg 3L^2/v_Fl=
\hbar/E_c$, the time of diffusion across the sample.

Let us discuss the possibilities of experimental verification of
the results. Although the $\Phi_0-$periodic term in the charge
fluctuations may not be easy to access directly, it leads to
observable effects in the common electric noise by making it
phase sensitive. The phase sensitivity was noticed recently and
studied in a different setup\cite{11} with the flux varying
periodically in time, $\Phi(t)= \Phi_a \cos\Omega t$, and
constant voltage $V$ applied across the conductor. It was found
that the noise as function of $V$ at $T\ll eV,\hbar\Omega$ has
cusp-like singularities at $V=n\hbar\Omega/e$ with the strengths
given by oscillating functions of the ac flux amplitude. The
oscillations and peculiar character of the limit
$\Omega\rightarrow0$ indicate the "almost static" nature of the
phase sensitivity which was named {\it non-stationary
Aharonov-Bohm effect}. Let us draw a connection of this effect
with the $\Phi_0-$periodic term in (\ref{result}). In the setup
of Ref.\cite{11} one has the flux $\Phi(t)=-cVt+\Phi_a\cos\Omega
t$ applied during all measurement time $2t_0$. Therefore,
  \begin{equation}\label{8}{\big
|}\int_{-t_0}^{t_0}e^{i\phi(t)+i\omega t}dt{\big |}^2=
\sum_nJ^2_n(2\pi\Phi_a/\Phi_0)4\pi
t_0\delta(\omega+n\Omega-eV/\hbar)) \ .\end{equation}
  Integrated over $|\omega|{d\omega\over2\pi}$ Expr.(\ref{8})
gives charge fluctuations as a sum of cusp-like singularities:
   \begin{equation}\label{9}\langle\!\langle
Q^2\rangle\!\rangle= 2t_0\sum_n\lambda_n|eV-n\hbar\Omega|\ ,
\end{equation}
    where $\lambda_n={{\rm g} e^2\over h} G_1 J^2_n(2\pi
\Phi_a/\Phi_0)$. To relate Expr.(\ref{9}) with the low frequency
noise spectral density of Ref.(\cite{11})  one divides it by a factor $2\pi
t_0$.
Comparison with Expr.(\ref{result}) makes it clear that the
singularities at integer $eV/\hbar\Omega$ are caused by the
logarithmic divergence of Expr.(\ref{result}), while the
oscillating dependence on $\Phi_a$ is related with the periodic
dependence of the first term of Expr.(\ref{result}) on $\Phi$.
Thus both features, singularities and oscillations, arise as
elements of the same phenomenon. The property of the singularities
(\ref{9}) to persist as $\Omega\to0$ is explained by weak
logarithmic dependence of Expr.(\ref{result}) on the duration
$\tau$ of the flux switching.

Before we close, let us mention another experimental situation
where the phenomena described above may lead to observable
effects. The possibility to change flux by a non-integer amount
is naturally realized in superconductors, since superconducting
flux quantum is ${1\over 2}\Phi_0$. Therefore, if a normal
conductor is surrounding a superconducting system in which the
flux is switching by ${n\over 2}\Phi_0$, e.g., due to quantum
tunneling, it will distinguish between odd and even $n$ by noise
level. In the case when Ohmic dissipation in normal metal is
strong, it may significantly suppress the odd flux tunneling
rate and lead to tunneling of flux pairs.

In conclusion, we studied charge fluctuations in a system driven
by flux varying from one constant value to another. In contrast
with the average transmitted charge proportional to the flux
change, the mean square of charge fluctuations contains a term
distinguishing between integer and non-integer flux change. As a
result, the dependence of noise on the flux is non-monotonous
and has minima near integer values. The special role of integer
flux is put in connection with the binomial statistics picture
of charge fluctuations, where the flux quanta are naturally
interpreted as discrete attempts to transmit charge.

\acknowledgements
We are grateful to Gordey Lesovik for the influence he had on
this work. Research of L.L. is partly supported by Alfred Sloan
fellowship.

\begin{figure}
\caption{ Mesoscopic disordered conductor with ideal leads bent in a loop
of  length  $L$  through  which  magnetic  flux $\Phi(t)$  is applied.}
\label{figure1}
\end{figure}


\begin{references}
\bibitem[*]{*} also at: L.D.Landau Institute for Theoretical Physics,
2, Kosygin str., Moscow 117334, Russia
\bibitem{Anderson}
P. W. Anderson, Phys.Rev.Lett.,{\bf 18}, 1049 (1967)
\bibitem{Nozieres}
P. Nozi\`eres, C. T. deDominicis, Phys.Rev., {\bf 178}, 1084 (1969)
\bibitem{Mahan}
G. D. Mahan, Many Particle Physics, Sec. 8.3,
(2-nd edition, Plenum Press, 1990)
\bibitem{1}
L. D. Landau and E. M. Lifshitz, Statistical Physics, Part 1,
\S 125 (Pergamon Press, Oxford, 1980)
\bibitem{2}
G. B. Lesovik, JETP Letters, {\bf 49}, p.594 (1989)
\bibitem{3}
B. Yurke and G. P. Kochanski, Phys.Rev.{\bf 41}, p.8184
(1989)
\bibitem{4}
M. B{\"u}ttiker, Phys.Rev.Lett., {\bf 65}, p.2901 (1990);
\bibitem{5}
S.-R. Eric Yang, Solid State Commun.{\bf 81}, 375 (1992)
\bibitem{6}
C. W. J. Beenakker, M. B{\"u}ttiker, Phys.Rev.B46, 1889 (1992)
\bibitem{7}
Th. Martin and R. Landauer, Phys.Rev.{\bf B45}, 1742 (1992)
\bibitem{8}
L. S. Levitov, G. B. Lesovik, JETP Letters {\bf 58}
(3), p.230, (1993)
\bibitem{9}
R. Landauer,  in: Localization,  Interaction   and
Transport Phenomena, eds. B.Kramer, G.Bergmann and Y.Bruynsraede
(Springer, Heidelberg, 1985) Vol.{\bf 61}, p.38;\\
see also a review by R. Landauer in: W.van  Haeringen  and
D.Lenstra  (eds.),  Analogies  in  Optics and Micro Electronics,
pp.243-257, Kluwer Academic Publishers (1990)
\bibitem{11}
G. B. Lesovik, L. S. Levitov, Noise in an AC biased junction.
Non-stationary Aharonov-Bohm effect. (preprint 9311048 at
cond-mat@babbage.sissa.it)
\end{references}
\end{document}